\documentclass[twocolumn,appendixfloats,tighten]{aastex6}
\usepackage{newtxtext,newtxmath}
\usepackage[T1]{fontenc}
\usepackage{ae,aecompl}
\usepackage{amsmath}	
\usepackage{amssymb}	
\usepackage{mathtools}
\usepackage{ctable}
\usepackage{url}
\usepackage{xspace}
\usepackage[normalem]{ulem}
\usepackage{hyperref}
\usepackage[all]{hypcap}
\usepackage{graphicx}
\usepackage{color}
\usepackage{float}

\def\msun{{\rm\,M_\odot}}

\def\msun{{\rm\,M_\odot}}

\newcommand{\be}{\begin{equation}}
\newcommand{\ee}{\end{equation}}

\newcommand{\chieff}{\ensuremath{\chi_{\rm eff}}\,}

\def\h2{${\rm\,H_2}$}

\usepackage{color}

\begin{document}

\title{The branching ratio of LIGO binary black holes }

\email{msafarzadeh@cfa.harvard.edu}

\author{Mohammadtaher Safarzadeh}
\affiliation{Center for Astrophysics | Harvard \& Smithsonian, 60 Garden Street, Cambridge, MA, USA\\ Department of Astronomy and Astrophysics, University of California, Santa Cruz, CA 95064, USA}

\begin{abstract} 
Formation of binary black holes (BBHs) detected by gravitational-wave (GW) observations could be broadly divided into two categories: 
those formed through field binary evolution and those assembled dynamically in dense stellar systems. 
The branching ratio of the BBHs refers to the contribution of each channel. 
The dynamical assembly channel would predict a symmetric distribution in the effective spins of the BBHs while field formation predicts BBHs to have positive effective spins. 
By modeling these two populations based on their effective spin distribution we show that in the 10 BBHs detected by LIGO/Virgo the contribution of the dynamically assembled BBHs to be more than about 50\% with 90\% confidence. 
This result is based on the assumption that the field binaries are born with positive effective spins not restricted to have small values.   

 \end{abstract}

\section{Introduction}

The effective spin of a binary black hole (BBH) system is defined as
\be
\chi_{\rm eff}\equiv\frac{m_1 a_1 \cos(\theta_1) +m_2  a_2  \cos(\theta_2)}{m_1+m_2} ,
\ee
 where $m_1$, and $m_2$ are the masses of the primary and secondary black hole, and $a_1$, and $a_2$ their associated dimensionless spin magnitude defined as:
 \be
 a=\frac{c J_{\rm BH}}{G M_{\rm BH}}.
 \ee
Here $c$ is the speed of light, G is the gravitational constant, and $M_{\rm BH}$ and $J_{\rm BH}$ are  the mass and angular momentum of the BH.
$\theta$ is the angle between the direction of each BH's spin and the orbital angular momentum of the BBH. 
The effective spin parameter is the best-measured spin-related parameter from gravitational wave observations \citep[][and references therein]{2017Natur.548..426F}.

The expected spin of a newly born BH depends on the efficiency of angular momentum (AM) transfer from the core of its progenitor star's core
to outer shell layers through magnetic fields. 
Models assuming moderate efficiency of AM transport through meridional currents predict the formation of BHs with high spins \citep{Eggenberger:2007dl,Ekstrom:2011ke}, 
while efficient transport by the Tayler-Spruit magnetic dynamo \citep{Spruit:1999vt,Spruit:2001ki}, as implemented in stellar evolution
calculations \citep{Fuller:2019gc,Fuller:2019jz} predicts all BHs to be born very slowly rotating. 

Therefore, the effective spin distribution of the binary black holes observed with LIGO/Virgo (hereafter, LIGO BBHs) illuminates their formation process \citep{2017CQGra..34cLT01V,2017Natur.548..426F,2017MNRAS.471.2801S}. 
Broadly, LIGO BBHs maybe divided into two categories: (i)
assembled in the field through stellar evolution and a potential common
envelope phase. Such binaries are expected to have their BH
spins preferentially aligned with the orbital angular momentum of the binary
\citep{Belczynski:2002gi,Dominik:2012cwa,Zaldarriaga:2017fn,Gerosa:2018hw,Qin:2018cl,Bavera:2019ut,2018ApJ...862L...3S}. 
(ii) assembled dynamically, either in globular or nuclear star clusters or hierarchical triple or higher-order stellar systems
\citep{Zwart:2004jj,2014ApJ...784...71S,Chatterjee:2016fl,Rodriguez:2016hi,Antonini:2017el,2018ApJ...853..140S,Rodriguez:2018ci}. 
Such binaries are expected to have their spin isotropically distributed with respect to the angular
momentum of the binary and therefore result in the symmetric distribution in \chieff.

While the effective spin parameter for the 10 LIGO/Virgo GWTC-1 BBHs is consistent with being clustered around zero \citep{Abbottetal:2018vb,Belczynski:2017wa,Roulet:2019js} 
recent work by \citet{2020arXiv200106490S} indicates a trend in the distribution suggestive of a non-negligible contribution from dynamically assembled binaries. 
In this \emph{Letter} we analyze the same set of BBHs searching for the contribution of field binaries.

\section{Method}\label{sec:method}
\begin{figure}
\includegraphics[width=1.0\linewidth]{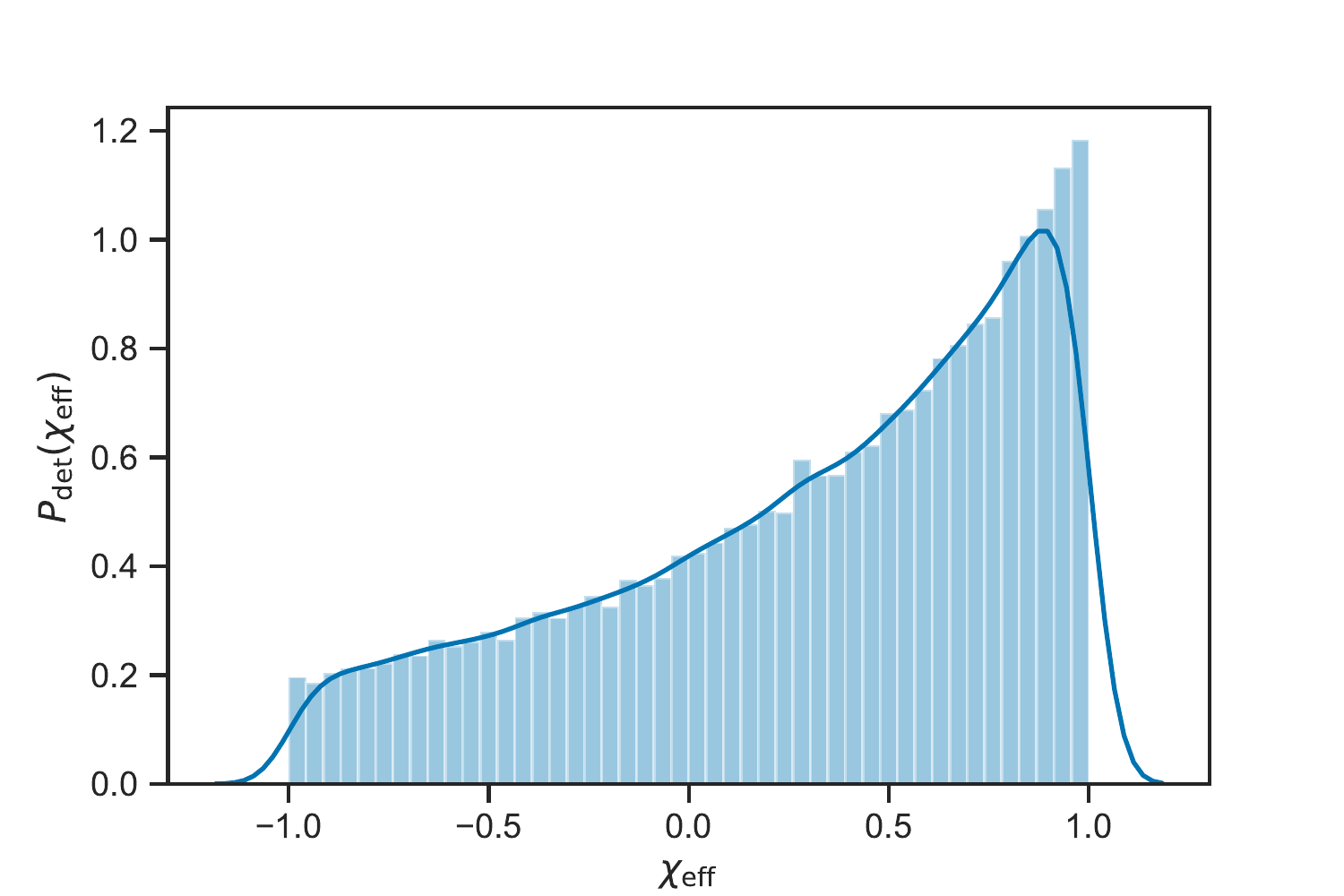}
\caption{Shows the detection probability for \chieff for a population of BBHs with $P(m_1)\propto m_1^{-1}$, and $p\left( m_2 \mid m_1 \right) \propto \mathrm{const}$ between 5 and 50 solar mass.
See \citet{2020arXiv200106490S} for details. }
\label{f:p_det}
\end{figure}

\begin{figure*}
\includegraphics[width=0.34\linewidth]{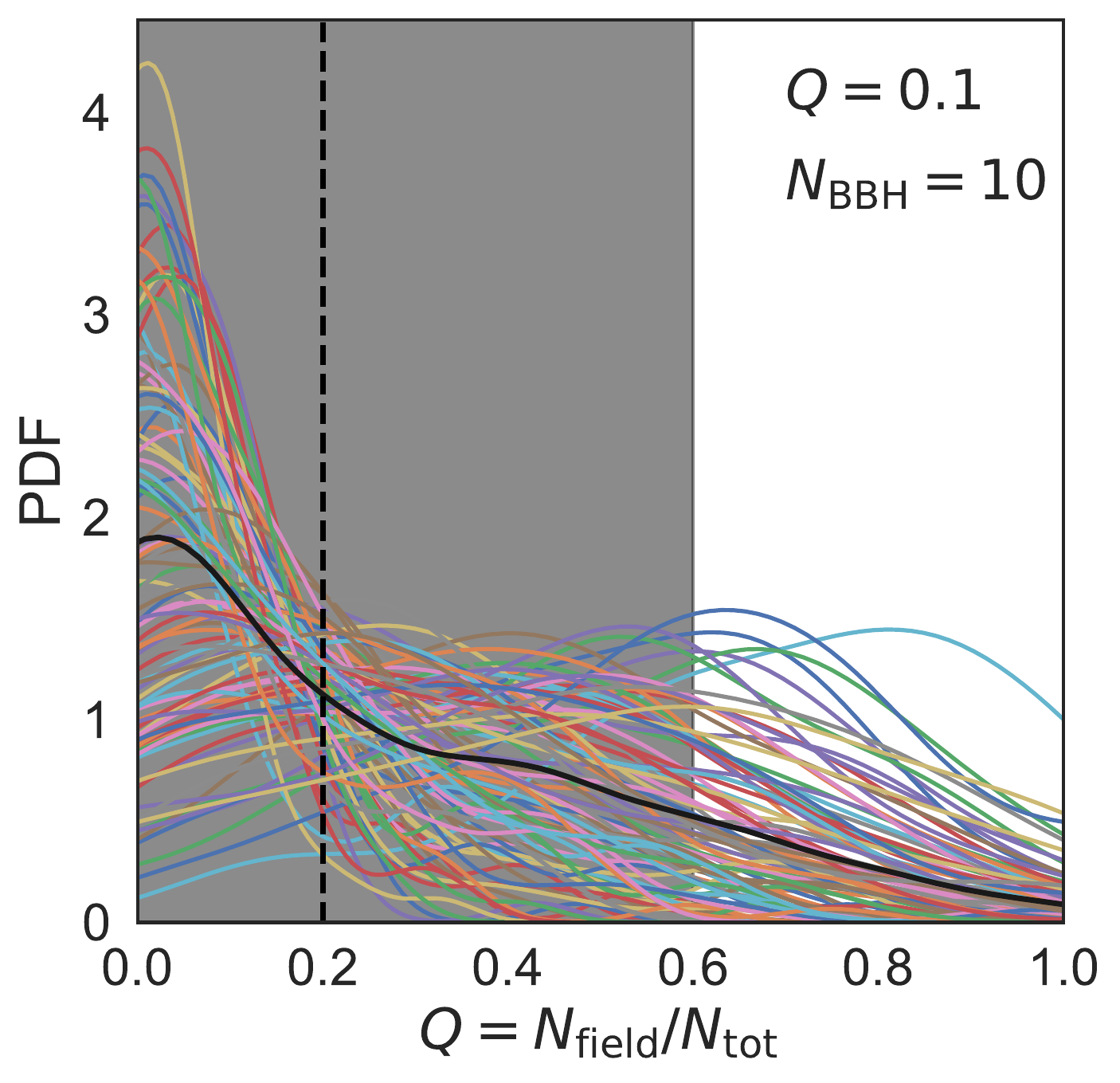}
\includegraphics[width=0.34\linewidth]{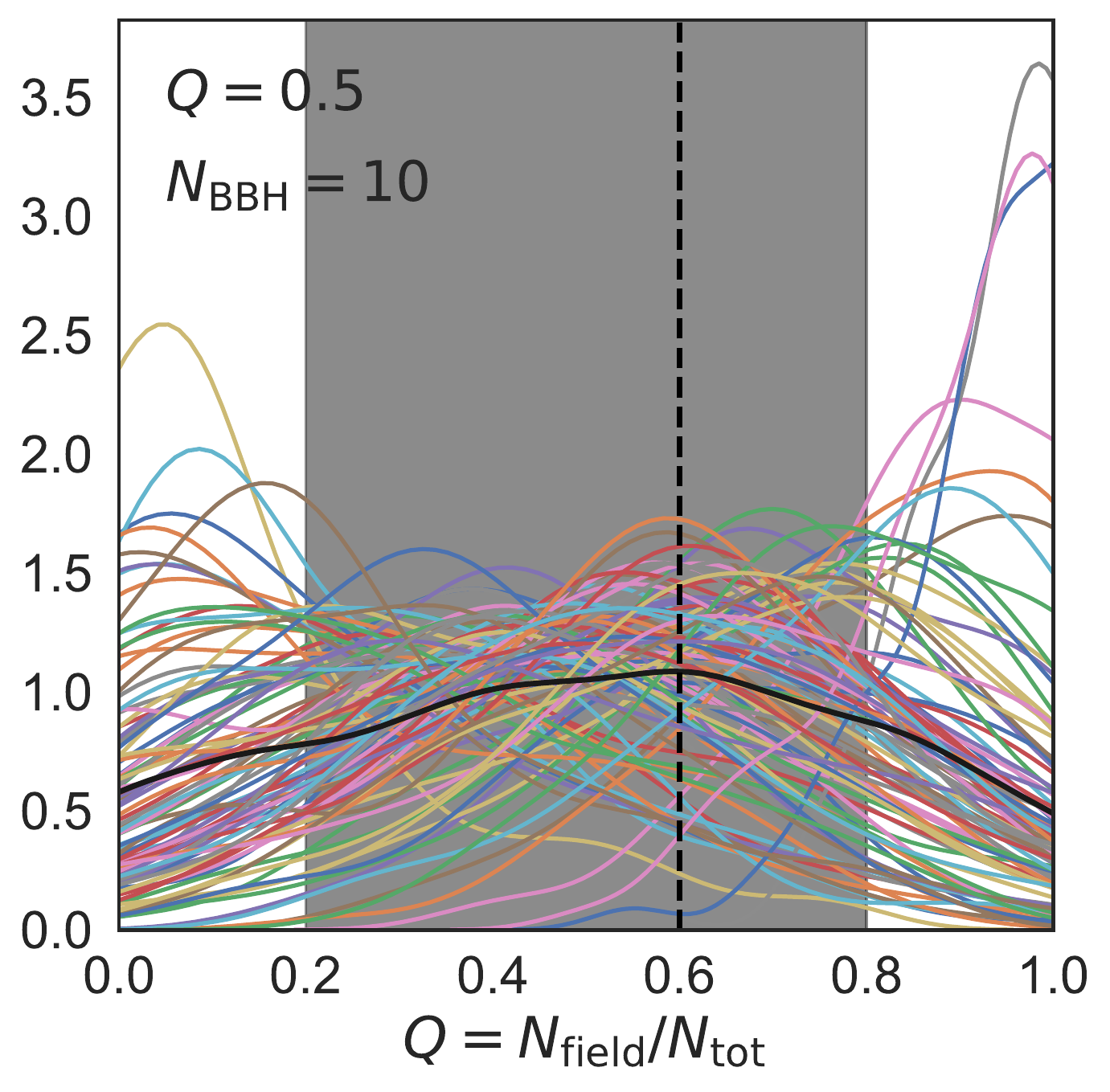}
\includegraphics[width=0.33\linewidth]{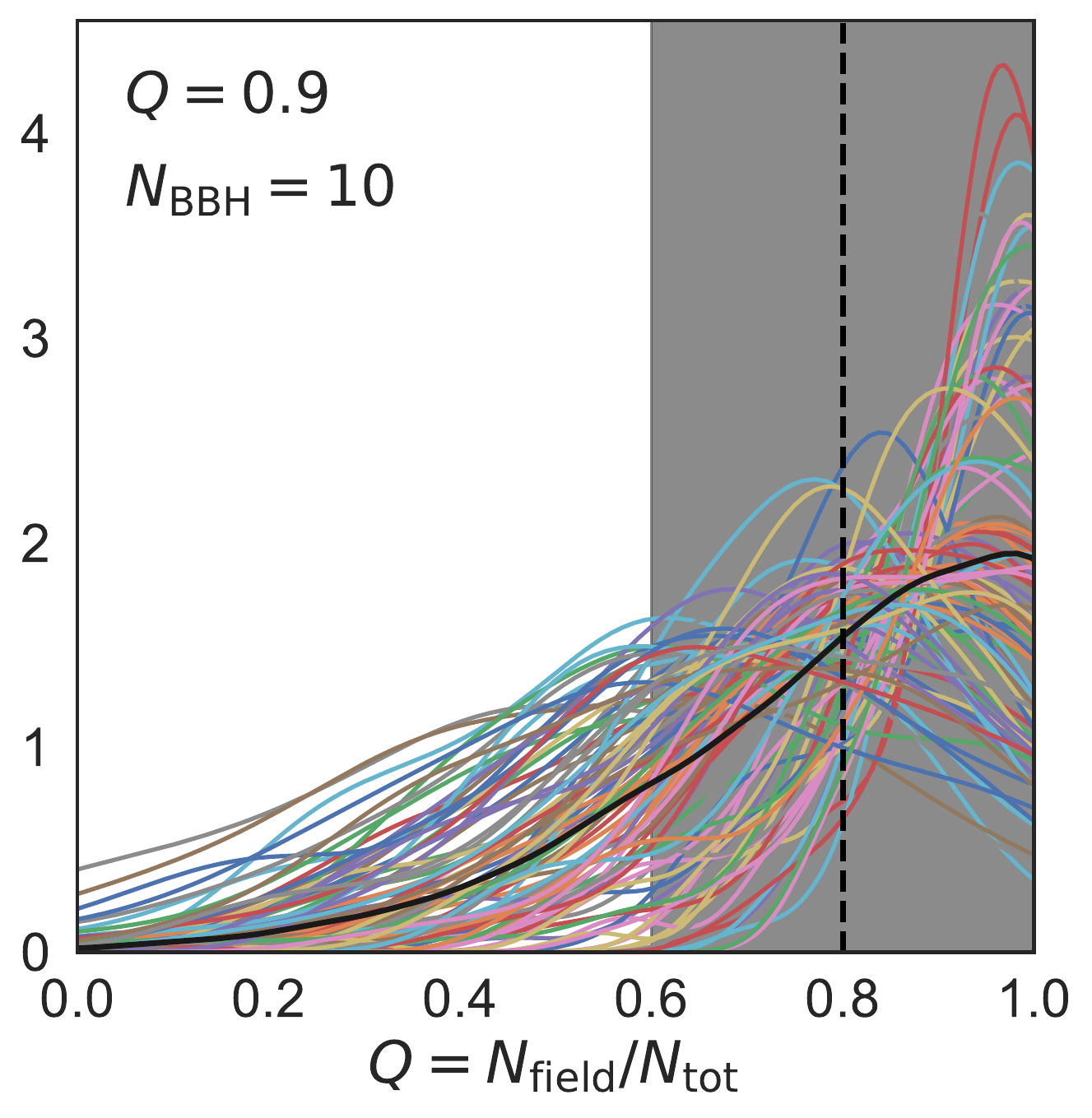}
\includegraphics[width=0.34\linewidth]{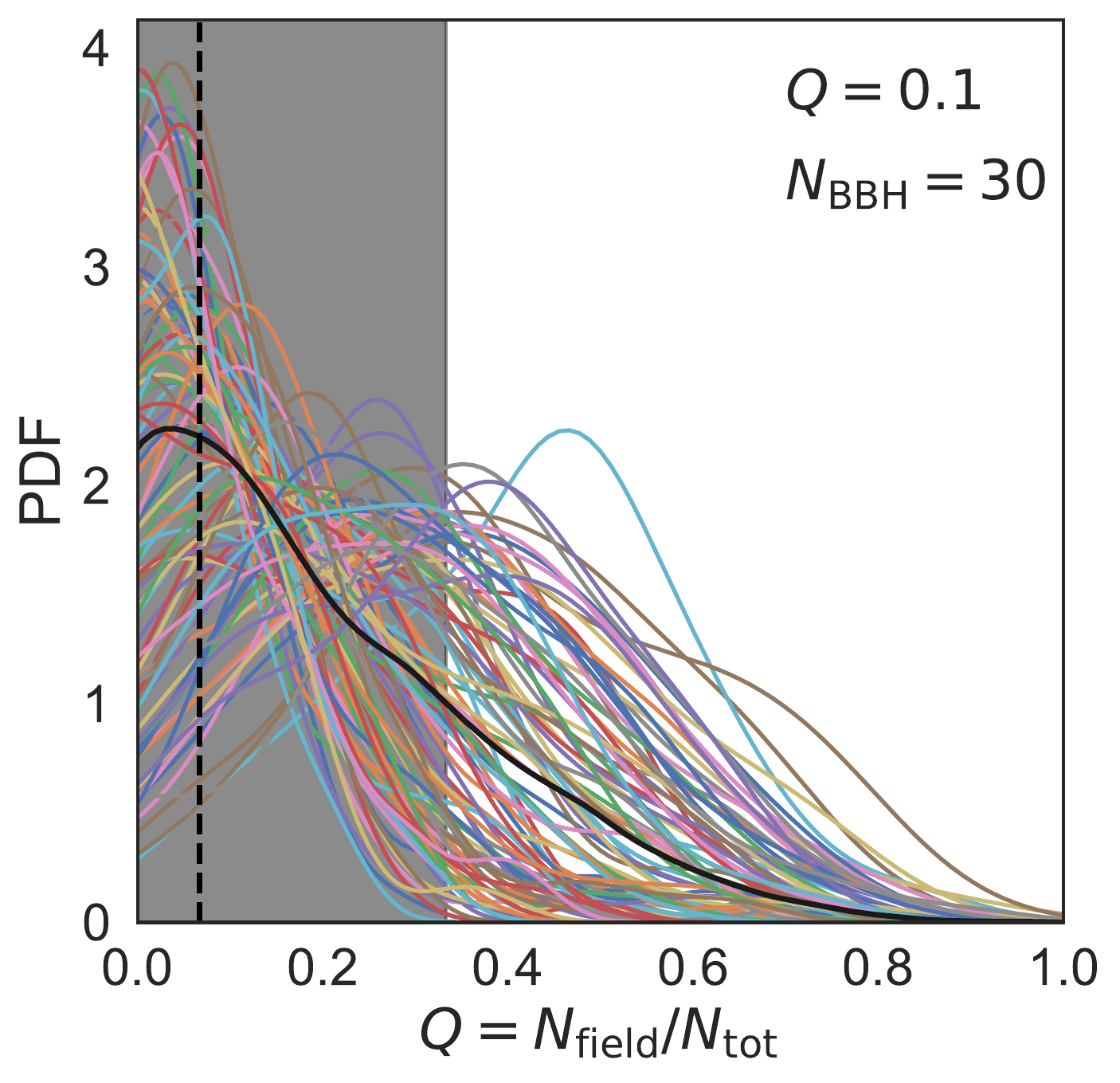}
\includegraphics[width=0.34\linewidth]{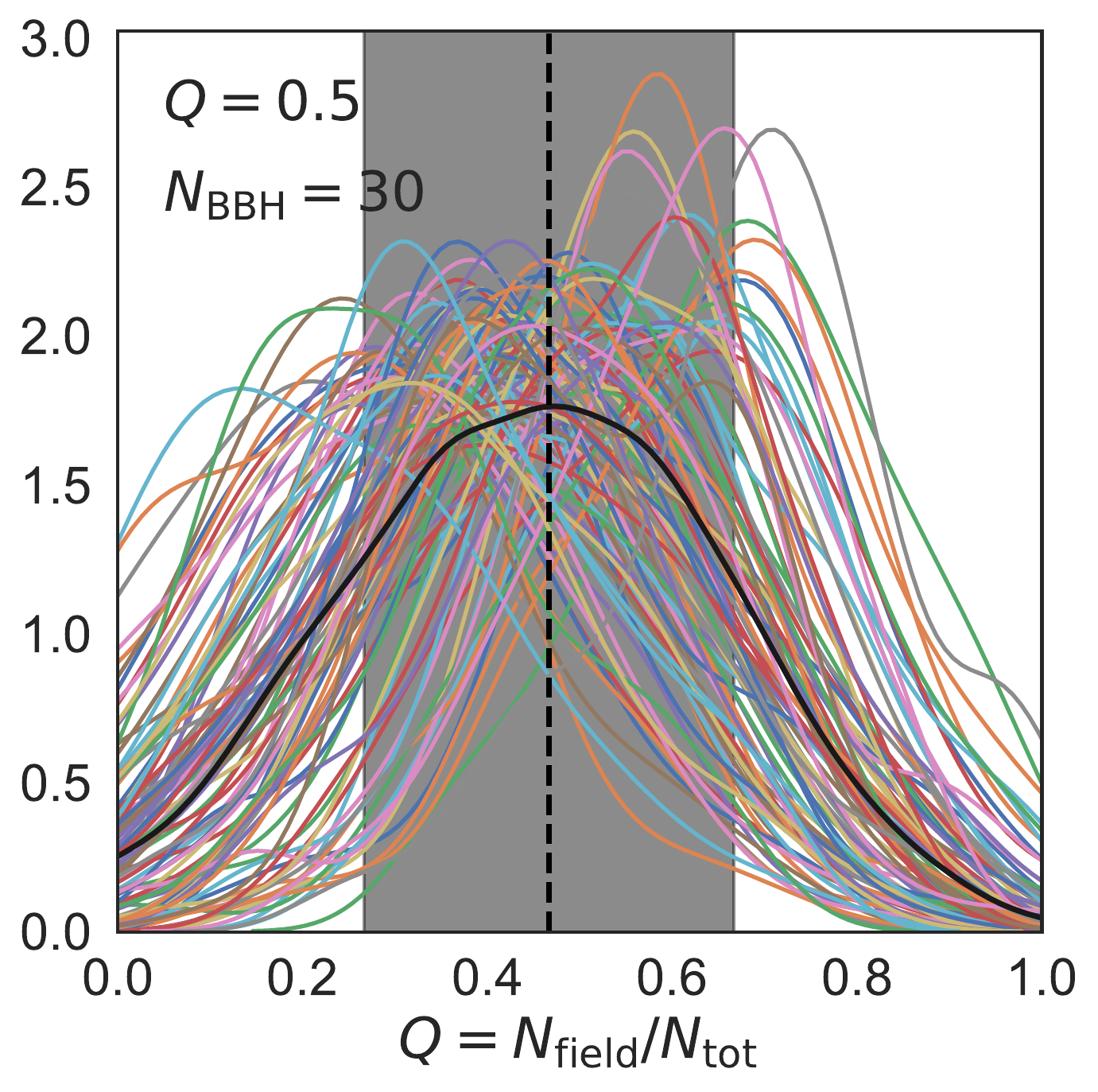}
\includegraphics[width=0.33\linewidth]{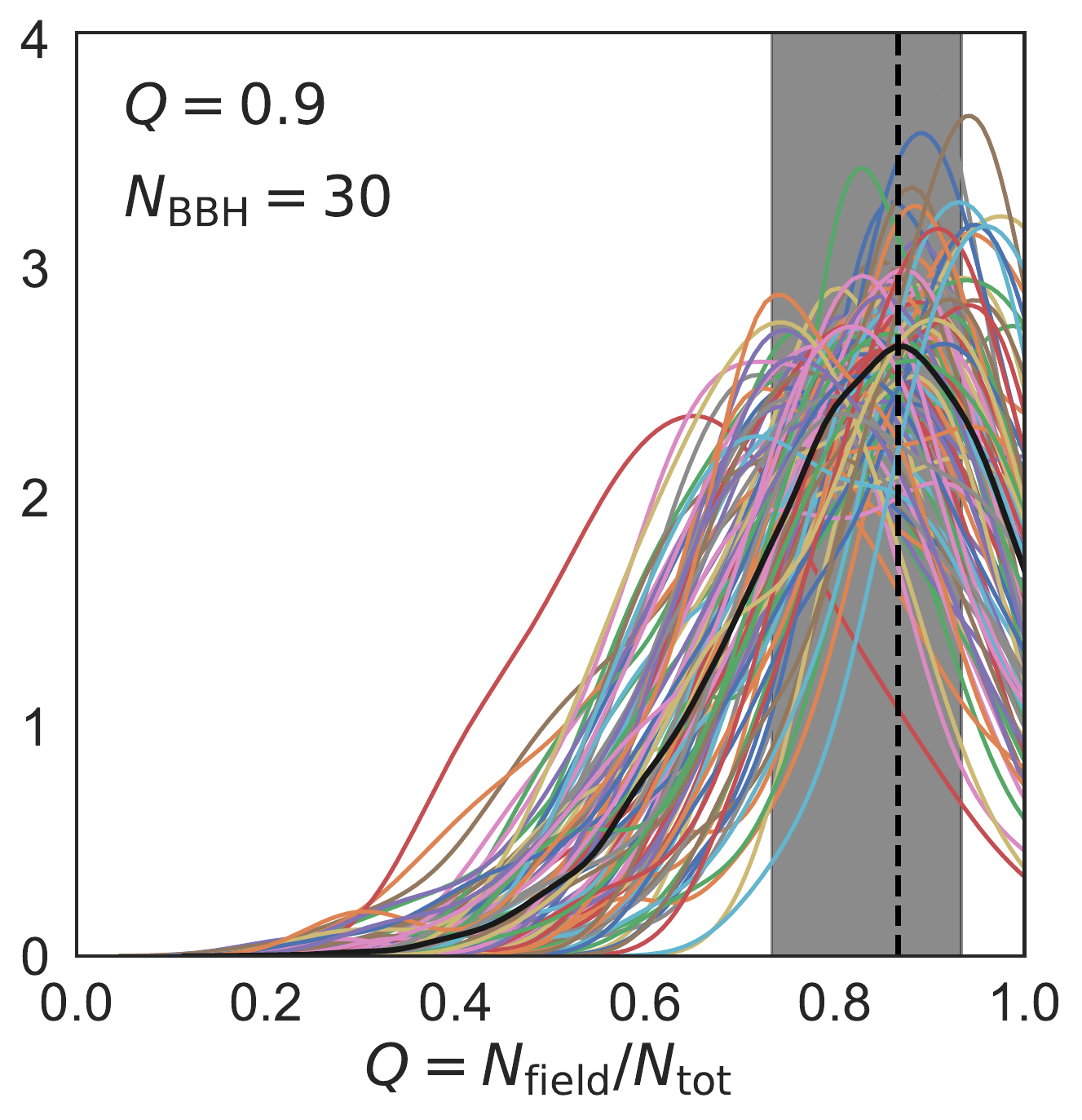}
\caption{\emph{Top row:} shows the posterior distribution on the parameter $Q$ which indicates the contribution fraction of field binaries to the overall mock BBH distributions.
Each colored line shows the posterior of $Q$ for a given random distribution of $N_{\rm BBH}=10$ BBHs in mass-\chieff plane. 
The black solid lines show the stack of the colored lines. The shaded region shows the 16th, and 84th percentile and the black dashed line shows the median value of the black PDF. 
Note that the dashed vertical black line does \emph{not} indicate the true $Q$ value.The true $Q$ is mentioned in each panel.
We note that in this computation the selection bias of LIGO in detecting BBHs with different \chieff is not modeled. \emph{Bottom row:} shows the same but when the $N_{\rm BBH}=30$. }
\label{f:mock}
\end{figure*}

The method is based on the assumption that dynamically assembled binaries will have symmetric distribution in \chieff while field binaries will prefer to have a predominantly positive \chieff distribution.
Although this assumption is broadly expected to be the case, there are mechanisms in the field formation scenario that can result in a negative effective spin \citep{Gerosa:2018hw} which we do not consider in the present work.
If we denote the number of BBHs with positive (negative) \chieff with $N_p$ ($N_n$), the fraction of BBHs coming from field population could be defined as:
\be
Q=\frac{N_p-N_n}{N_p+N_n}.
\ee

Although dynamical assembly of the BBHs would predict symmetric distribution in \chieff, LIGO's sensitivity for the detection of a BBH is not insensitive to their \chieff. 
Figure \ref{f:p_det} shows the detection probability of a population of BBHs with $P(m_1)\propto m_1^{-1}$, and $p\left( m_2 \mid m_1 \right) \propto \mathrm{const}$ and each component mass range between 5 and 50 $\msun$.
For LIGO it is easier to detect BBHs with positive \chieff than their negative counterparts \citep{2018PhRvD..98h3007N}, and therefore, even if all the BBHs are assembled dynamically, LIGO would be biased towards those with positive \chieff making the final 
distribution in mass-\chieff plane asymmetric. 

Therefore, the formula above would need to be corrected for the selection bias of LIGO. 
When analyzing LIGO BBHs we instead use the effective $N_p$, and $N_n$ defined as:
\be
<N_p>=\sum_{i=0}^{i=N_{BBH}} P_{\rm det}^{-1}(\chi^{i}_{\rm eff}) \mathcal{H}(\chi_{\rm eff}^i)
\ee
and 
\be
<N_n>=\sum_{i=0}^{i=N_{BBH}} P_{\rm det}^{-1}(\chi^{i}_{\rm eff}) \mathcal{H}(-\chi_{\rm eff}^i), 
\ee
where $\mathcal{H}(x)$ is Heaviside step function returning 1 if the argument is positive and zero otherwise.

If we have a set of $N$ BBHs, by sampling $N_s$ times from their posterior distribution in mass and \chieff we can have $N_s$ times measurement of the parameter $Q$ and construct a probability distribution for $Q$.
We adopt $N_s=100$ for testing purposes and $N_s=1000$ for analyzing LIGO data.
We do this by sampling with replacement to remove potential bias from individual BBHs in the sample, specifically important when the sample size is small. 

\begin{figure}
\includegraphics[width=\linewidth]{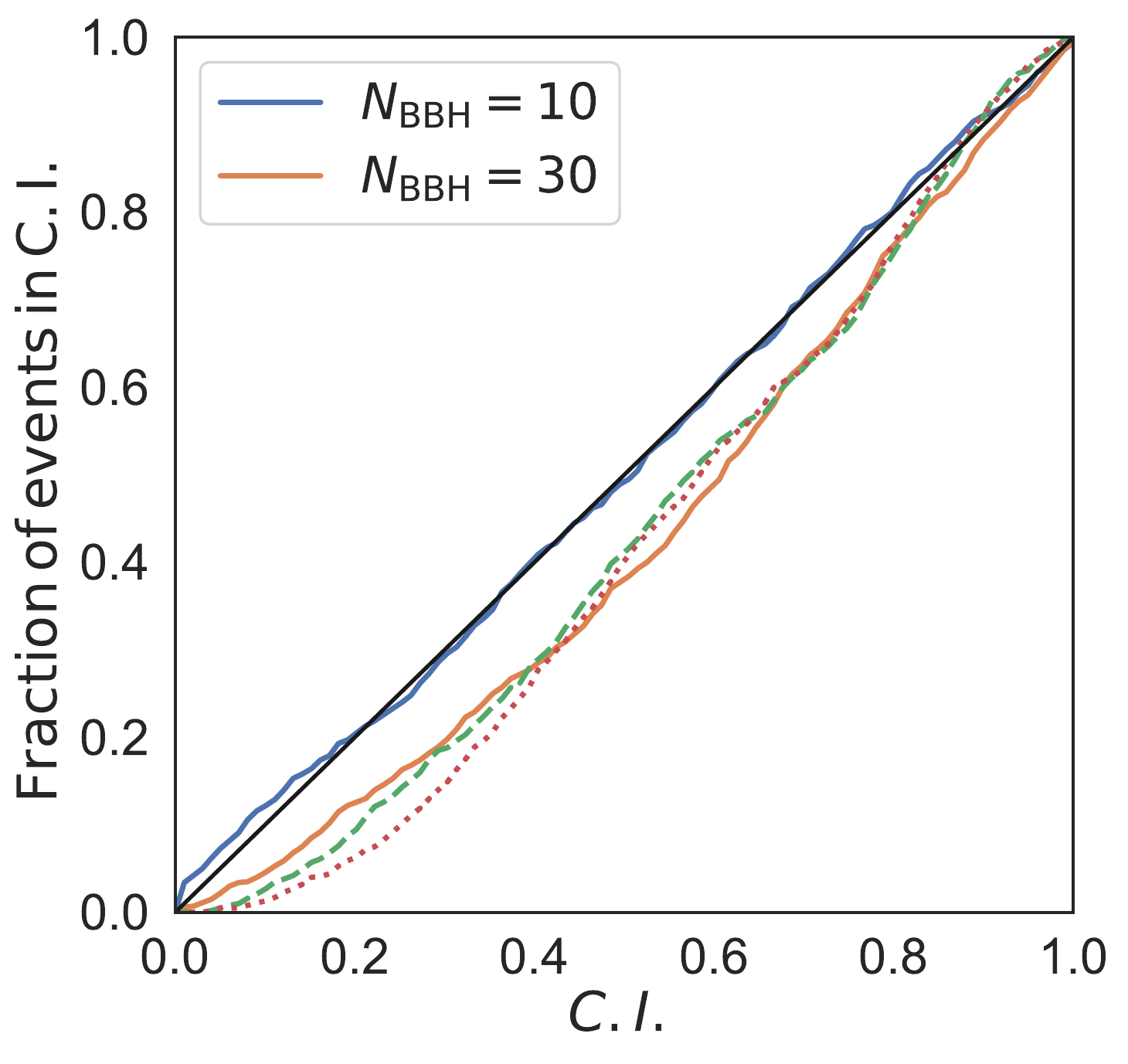}
\caption{Shows the fraction of simulated BBH distributions with Q values within a credible interval as a function of credible interval. 
If our parameter estimation method is unbiased, one expects to recover the black solid diagonal line which indicates the ideal 1-to-1 relation.
The blue line shows our results from 1000 simulation of a set of 10 BBHs with random underlying $Q$ values. This result indicates that on average we under estimate the true input $Q$. 
The dashed green (dotted red line) shows the test result when we impose a bias of 5\% (10\%), respectively. This shows that the bias in our method lies between 5-10\%. }
\label{f:pp}
\end{figure}

\begin{figure}
\includegraphics[width=\linewidth]{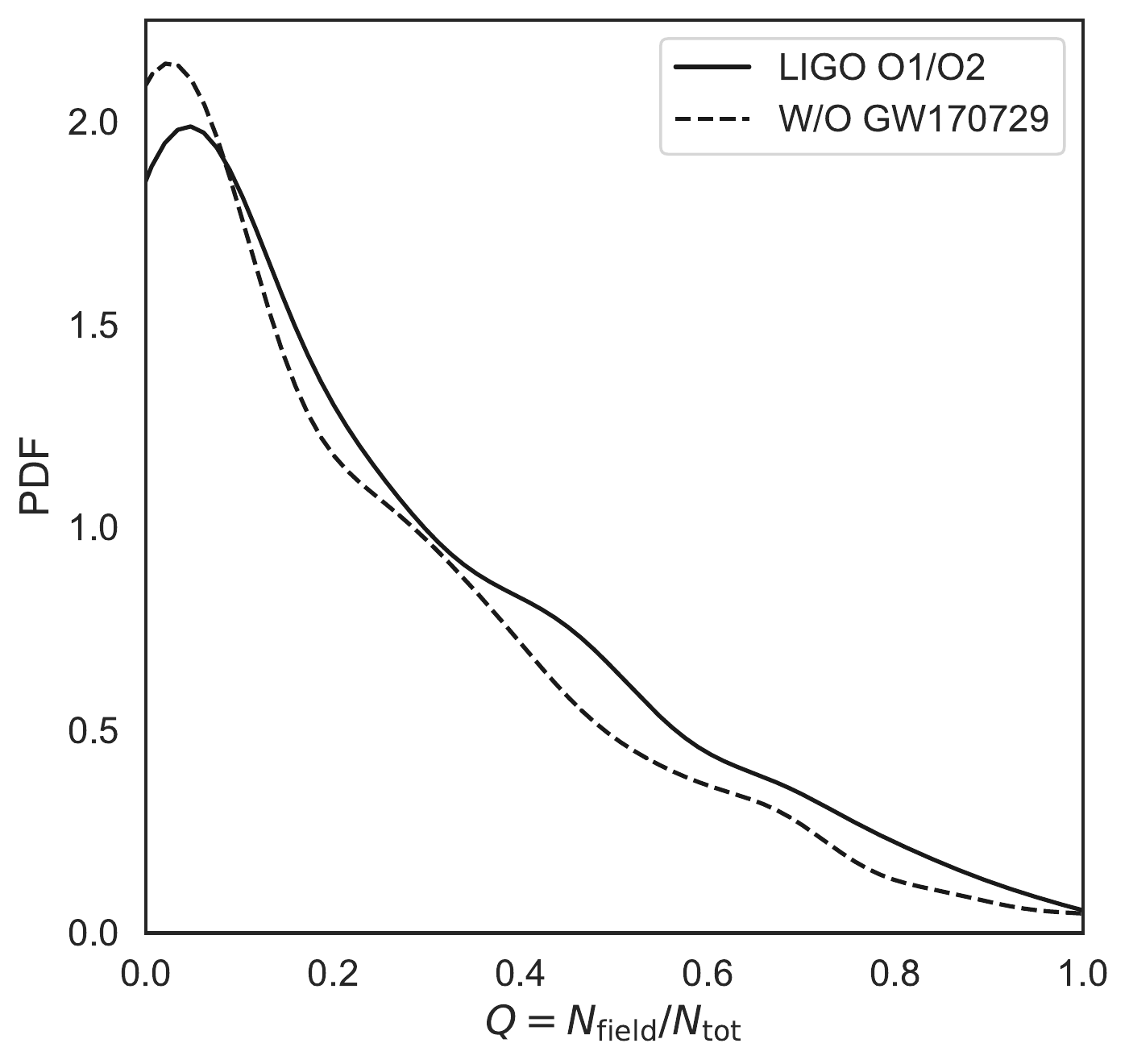}
\caption{Shows the posterior distribution on the parameter $Q$ which indicates the contribution fraction of field binaries to the overall BBH observed with LSC.
In solid black line we show the result for all the 10 BBHs, and in dashed black line we have excluded GW170729 from the analysis given this BBHs has the highest false alarm rate. 
Both lines are smoothed KDE representation of the results.
Our results indicate that the contribution from the dynamical channel is more than $\approx 55\%$ with 90\% confidence, and excluding GW170729 does not have a significant impact on the overall statistics.}
\label{f:ligo}
\end{figure}

To test the method, we first construct a set of mock distribution of BBHs in \chieff-mass plane. 
After defining the total number of the BBHs and $Q$, the total number of the field BBHs is set through a binomial distribution $N_f=\binom{N_{\rm BBH}}{Q}$, and the 
total number of dynamically assembled binaries is $N_d=N_{BBH}-N_f$. 
For all binaries we assume a uniform distribution in chirp mass between 10 and 50 $\msun$. 
For field binaries we assume a uniform distribution in \chieff$\in [0,1]$. For the dynamically assembled binaries we consider a normal distribution $\mathcal{N}(0,\sigma)$ with $\sigma= \alpha(M_c/5)$, where $M_c$ is the chirp mass of the binary in solar units.
Such a distribution would populate dynamically assembled binaries symmetrically in \chieff around zero with its dispersion increasing with mass. 
We set a fiducial value of $\alpha=0.1$, however, our result is not sensitive to the exact choice of this parameter. 
The increase of dispersion is due to a random walk in \chieff-mass that higher generation BHs follow.
After populating the BBHs in mass-\chieff plane, we assume their posterior probability distribution function (PDF) in mass and \chieff follows a normal distribution with $\sigma_{\chi_{\rm eff}}=0.1$, and $\sigma_{M}=1\msun$ which 
is similar to the dispersion in the posterior distribution on mass and \chieff of the first 10 LIGO BBHs \citep{Abbott_2019}. 
When sampling from the \chieff PDF of the BBHs, we impose a minimum and maximum of -1, and 1 for the \chieff.

Figure \ref{f:mock} shows three different mock distribution where their true $Q$ is shown with a dashed black line. 
In each panel, we simulate $N_{\rm BBH}$ mock BBH observations, and determine the recovered Q parameter. In the top (bottom) row we set $N_{\rm BBH}=10\,(30)$.
We repeat this process 100 times, and show each realization with a different color in Figure \ref{f:mock}.
The black solid line shows the stack of all the colored lines, indicating the overall power of recovering the true $Q$ when the number of the BBHs is considered to be 10. 
We note that in this computation the selection bias of LIGO in detecting BBHs with different \chieff is not modeled. 

The summary of our results is presented in the probability-probability plot (p-p plot) shown in Figure \ref{f:pp}. 
This plot shows the fraction of simulated BBH distributions with Q values within a credible interval as a function of credible interval. 
If our parameter estimation method is unbiased, one expects to recover the solid black diagonal line which indicates the ideal 1-to-1 relation. 
The blue curve shows our results. We have done the following steps: 
i) Draw $N_{\rm inj}=1000$ $Q$ values from a uniform prior between 0 and 1.
ii) Generate mock dataset : draw $N_{\rm BBH}$ BBHs from the mixture model described in section 2 with a true value of $Q$.
iii) Produce posterior on $Q$ for each of the $N_{\rm inj}$. 
iv) Determine at which credible interval the true injected value of Q falls within each mock posterior.
v) For each credible interval between 0 and 1, compute the fraction of events which contain the true value within that credible interval. 
The blue (orange) line in Figure \ref{f:pp} indicates the result when $N_{\rm BBH}=10\,(30)$. 
As can be seen the relationship we get for $N_{\rm BBH}=10$ is close to a 1-to-1 relation but the result for $N_{\rm BBH}=30$ deviates from it, and therefore our parameter estimation is biased in the sense that we tend to \emph{under estimate} $Q$.

The source of the bias lies in the magnitude of the error we consider for \chieff. Throughout this paper we have assumed $\sigma_{\chi_{\rm eff}}=0.1$ based on the LIGO GWTC-1 catalog.
When sampling from the posterior distribution of a BBH's \chieff, the sampled realization can have a negative \chieff value depending on the magnitude of the BBH's mean \chieff. 
Since in this formalism any BBH with negative \chieff is considered to be coming from a dynamical channel, we will be biased to under estimate the contribution of the field binaries. 
For example if we adopt a $\sigma_{\chi_{\rm eff}}=0.01$ for the BBHs, we would not see a bias in the results. The impact of this bias is more pronounced when the sample size is increased.
Moreover, if we assume that field binaries are all born with small positive effective spins (e.g., \chieff <0.1), results from our method would be greatly biased in favor of dynamically assembled binaries because of the 
current effective spin error magnitude considered in this work (i.e, $\sigma_{\chi_{\rm eff}}=0.1$).

In order to quantify the level of bias in our method, we perform the following test: i) Draw $N_{\rm inj}=1000$ $Q$ values from a uniform prior between 0 and 1.
ii) Produce posterior on $Q$ for each of the $N_{\rm inj}$. The posterior on $Q$, however, is assumed to be a normal distribution centered on $Q$ with standard deviation of $\sigma_t=0.5$.
iii) We re-assign the true value of $Q$ by drawing from the posterior we made in the previous step. 
iv) In order to make the result biased, we shift the true $Q$ value from the previous step by $\delta Q$. 
v) Determine at which credible interval the true injected value of Q falls within each mock posterior and plot the p-p plot.
We assume $\delta Q$ to be 0.05 (0.1) and the result is shown with dashed green (dotted red) line in Figure \ref{f:pp}.
As can be seen the orange line in Figure \ref{f:pp} lies between the results where we impose 5\% and 10\% bias in under estimating the truth.
We note that in the above test, we are not sensitive to the exact choice of $\sigma_t$. For example, adopting $\sigma_t=0.1$ (meaning a narrow PDF for the 
posterior on $Q$) would make our result to be consistent with a bias of less than 5\% while a larger value of $\sigma_t$ would not increase the bias above 10\%.

\section{GWTC-1 result}\label{sec:LIGO}

In this section, we present our result on the catalog of the first 10 BBHs observed by the LSC\footnote{See \url{https://dcc.ligo.org/LIGO-P1800370/public}} \citep{Abbott_2019}. 
Figure \ref{f:ligo} shows the posterior distribution for $Q$. In solid black line we show the result of taking into account the detection probability function of LIGO
as a function \chieff and analyzing all the ten LIGO BBHs. Our results indicate that the contribution from the dynamical channel is more than $\approx 55\%$ with 90\% confidence. 
In dashed black line we show the same result when excluding GW170729 from the analysis since this BBH merger event has the highest false alarm rate among all. We see that our result is not driven 
by GW170729 although excluding this event slightly increases the contribution of the dynamical channel to the overall statistics. 

We note that if the field binaries are all born with very small positive effective spin, the likelihood of considering them as dynamically assembled increases given the current level of uncertainty on effective spin $\sigma_{\chi_{\rm eff}}\approx0.1$. 
In this case our method would be biased in favor of dynamically assembled binaries unless the posterior that LIGO provides for such sources are  smaller than their mean effective spin magnitude. 

\section{Summary \& Conclusion}\label{sec:summary}
BBHs observed by LIGO/Virgo are expected to populate different areas in the \chieff-mass plane depending on their formation channel
Field binaries tend to predict a largely positive \chieff distribution while dynamical assembly of BBHs in dense stellar clusters leads to a symmetric distribution of BBHs in \chieff at all masses with larger dispersion at higher masses.
The increase of dispersion is due to a random walk in \chieff-mass that higher generation BHs follow.

In \citet{2020arXiv200106490S} we found a tentative negative correlation between \chieff and chirp
mass for the ten LIGO/Virgo BBHs with $\sim 75\%$ confidence. Moreover, we found that the dispersion in \chieff grows
with mass with  80\% confidence. These trends are consistent with a combined
channel of dynamically assembled BBHs that provide the positive trend of
dispersion with mass, and a field formation channel that provides the negative
mean trend with mass could explain our findings.

In this \emph{Letter} we took a different approach to characterize the branching ratio of the LIGO BBHs. 
The fundamental assumption in this work is that dynamically assembled BBHs 
will be distributed symmetrically in \chieff (prior to correcting for the LIGO detection bias of the BBHs as a function of their \chieff), 
while BBHs formed through field binary evolution will end up as having a positive \chieff although each sub-channel will populate 
a distinct region in \chieff-mass plane. We note that the analysis itself is only carried out in effective spin dimension and not in the \chieff-mass plane.

We show that there is a support for the symmetric distribution in \chieff (which is the tracer of the dynamical formation channel) to be more than 55\% with 90\% confidence. 
This result is not sensitive to the presence of GW170729 which is a particular high spin event in the catalog and is in line with the findings in \citet{2020arXiv200106490S}. 
By taking a rather different approach, while having the same goal as the work presented here, \citet{Farr_2018} analyzed the first four LIGO BBHs and concluded that the odds of them being 
formed from field distribution over dynamically assembled origin is 1.1. Implementing a hierarchical Bayesian approach \citep{Abbottetal:2018vb} show that the data prefers a dynamical assembly origin for the LIGO BBHs, however, 
assuming the intrinsic effective spins are clustered around zero would significantly reduce our ability to distinguish between the two formation channels.
Explaining the LIGO BBHs with the latest catalogs of population synthesis models and N-body codes, \citet{2019ApJ...886...25B} 
concludes that the data is barely consistent (still consistent) with a model in which all the BBHs are born in the field (clusters), a result that would depend on the metallicity distribution of the BBH progenitors \citep{2019ApJ...883L..24S}.

Our approach in this work is simple with basic assumptions about effective spin distribution of the BBHs. We show that such simple assumptions leads to conclusions that are consistent with previous Bayesian modeling of the 10 LIGO BBHs.
Other consideration can in principle be incorporated in such analysis: for example if the LSC detects a BBH with mass above the pair-instability mass gap \citep{Woosley:2017dj}, 
the likelihood of such BBH to belong to dynamical formation channel would be increased. 
Likewise high mass ratio systems are likely formed in the field since such systems can not be effectively assembled in clusters due to mass segregation (see \citet{2019arXiv191104495S} and references therein).
Future O3 data release will confirm this finding as the number of events is expected to be around 30, which increases the overall sensitivity by about a factor of 2. 

\software{
Numpy \citep{numpy},
Scipy \citep{scipy},
IPython \citep{IPython},
Matplotlib \citep{Matplotlib}
}
\acknowledgements
MTS is thankful to the referee their constructive comments. MTS is thankful to Avi Loeb for asking me about the branching ratio of the LIGO BBHs which inspired this work.
MTS is also thankful to Will M. Farr, Enrico Ramirez-Ruiz, and Ryan Foley for insightful discussions.  
MTS thanks the Heising-Simons Foundation, the Danish National Research Foundation (DNRF132) and
NSF (AST-1911206 and AST-1852393) for support. 

\bibliographystyle{apj}
\bibliography{the_entire_lib.bib}

\end{document}